# Formal Concept Analysis for Information Retrieval

Abderrahim El Qadi*,‡

*Department of Computer Science
EST, University of Moulay Ismaïl,
Meknes, Morocco
elqadi_a@yahoo.com

Driss Aboutajdine‡

‡GSCM-LRIT, Unité associée au
CNRST, URAC 29, Faculty of Science,
University of Mohammed V,
Rabat-Agdal, Morocco

Yassine Ennouary‡

‡GSCM-LRIT, Unité associée au CNRST,
URAC 29, Faculty of Science, University
of Mohammed V,
Rabat-Agdal, Morocco

*Abstract*—In this paper we describe a mechanism to improve Information Retrieval (IR) on the web. The method is based on Formal Concepts Analysis (FCA) that it is makes semantical relations during the queries, and allows a reorganizing, in the shape of a lattice of concepts, the answers provided by a search engine. We proposed for the IR an incremental algorithm based on Galois lattice. This algorithm allows a formal clustering of the data sources, and the results which it turns over are classified by order of relevance. The control of relevance is exploited in clustering, we improved the result by using ontology in field of image processing, and reformulating the user queries which make it possible to give more relevant documents.

## I. INTRODUCTION

The World Wide Web (WWW) has become the most popular information source for people today. One of the major problems to be solved is related to the efficient access to this information that is retrieved by human actors or robots (agents). Our work falls under this context. We propose a solution to seek the relevant sources within sight of a query user. The data sources which we consider are the research tasks of laboratory LRIT[1] of the Faculty of Science Rabat Morocco. Facing such a problem, we seek in this work to analyze more precisely inter-connected themes between the authors, the publications and sets of themes of LRIT laboratory.

There was some interest in the use of lattices for information retrieval by [1, 2]. These systems build the concept lattice associated with a document/term relation and then employ various methods to access the relevant information, including the possibility for the user to search only those terms that has specified. Building the Galois (concept) lattice can be considered as a conceptual clustering method since it results in a concept hierarchy [3, 4]. This form of clustering constitutes one of the motivations of the concept's application lattice for IR. This comes owing to the fact that clustering out of lattice makes it possible to combine research by query and research by navigation.

Consequently the concept lattice generated from unit objects represents, in an exhaustive way, the possible clustering between these objects. Each cluster corresponding to a concept. Some of these concepts bring redundant information and are less interesting. This redundancy of information due to is made that the properties are showed as independent, and the possible existence of the semantic relations between the properties is not taken into account. On the other hand, the semantic relations between the properties can exist. So it proves to be useful to use ontology or taxonomy of field. In order to make correspond as well as possible the relevance of the user and the relevance of the system, we used a stage of query reformulation. The initial query is treated like a test to find information. The documents initially presented are examined and a formulation improved of the query is built from ontology, in hope to find more relevant documents. The query reformulation is done in two principal stages: to find terms; extension to the initial query, and to add these terms in the new query.

The paper is organized as follows: Section 2 introduces the Ontology (taxonomy), and in section 3 we presented the kinds of query reformulation used. In section 4 we illustrate FCA. Section 5, report the procedures and describe the system implemented for building concepts lattice and IR, and we show the results obtained. Section 6 offers some conclusions related to this work.

## II. ONTOLOGY

The concept of ontology became a key component in a whole range of application calling upon knowledge. Ontology is defined like the conceptualization of the objects recognized like existing in a field, their properties and relations connecting them. Their structure makes it possible to represent knowledge of a field under a data-processing format in order to make them usable for various applications.

An ontology can be constructed in two ways: domain-dependent or generic. Generic ontologies are definitions of concepts in general; such as WordNet [5], which defines the meaning and interrelationships of English words. A domain-dependent ontology generally provides concepts in a specific domain, which focuses on the knowledge in the limited area, while generic ontologies provide concepts more comprehensively.

The implementation of ontology is generally taxonomy of concepts and corresponding relations [6]. In ontology, concepts are the fundamental units for specification, and provide a foundation for information description. In general, each concept has three basic components: terms, attributes and relations. Terms are the names used to refer to a specific

---

concept, and can include a set of synonyms that specify the same concepts. Attributes are features of a concept that describe the concept in more detail. Finally relations are used to represent relationships among different concepts and to provide a general structure to the ontology. Figure 1 is an example of a simple ontology about the organization of concepts used in image processing.

In this ontology example, every node is a concept defined in image processing field. For each concept, there should be a set of attributes used to specify the corresponding concept. For instance, for the concept "Segmentation", the attributes of name and type are shown, and help explain the corresponding concept.

The relations between different concepts are also simplified. In a real application, several types of concept relations are used.

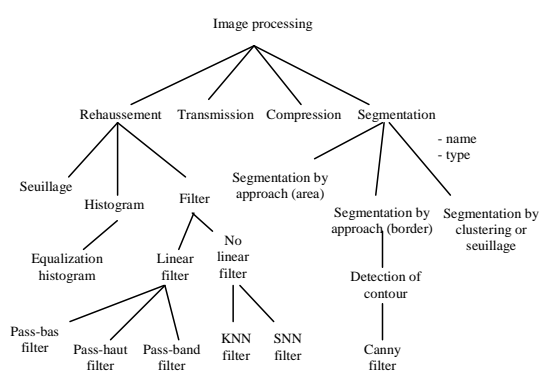

Figure 1.   An example of ontology in field of image processing

## III.   QUERY REFORMULATION

However, it is often difficult for user to formulate exact his requirement in information. Consequently, the results which the SRI provides them are not appropriate. To find relevant information by using the only initial query is always difficult, and this because of inaccuracy of the query. In order to make correspond as well as possible the relevance of the user and the relevance of the system, a stage of query reformulation is often used.

The query reformulation can be interactive or automatic [7]. The interactive query reformulation is the strategy of reformulation of the most popular query. It is named commonly re-injection of the relevance or "relevance feedback". In a cycle of re-injection of relevance, one presents to user a list of documents considered to be relevant by the system like answer to the initial query. After examined then, user indicates how he considers them relevant. This system allows users to expand or refine their query through the use of relevance feedback [8]. The typical scenario begins with a user indicating which documents retrieved from a query are most relevant. The system then tries to extract terms which co-exist in these documents and adds them to the original query to retrieve more documents. This process can be repeated as many times as desired. However, the limitation of this approach is that users are often required to place a bound on the number of documents retrieved as their query may be too general, and hence, retrieve too many irrelevant documents.

An alternative approach that has gained in interest recently is to apply the FCAs [9]. The advantage of this approach is that users can refine their query by browsing through well defined clusters in the form of a graph. The principal idea of the re-injection of relevance is to select the important terms belonging to the documents considered to be relevant by user, and to reinforce importance of these terms in the new query formulation. This method has double advantage a simplicity run for user who s' doesn't occupy of the details of reformulation, and a better control of the process of research by increasing the weight of the important terms and by decreasing that of the non important terms. In the case of automatic reformulation, user does' not intervene. Extension of the query can be carried out to leave a thesaurus or an ontology, which defines the relations between the various terms and makes it possible to select new terms to be added to the initial query.

In this work, to hold account the semantic relations between the concepts; we used an ontology presented in figure 1. This ontology will be used for query reformulation, which we used two types of modes respectively reflect reformulation by generalization and reformulation by specialization:

- Reformulation by generalization: consists in locating the top c of T (tree) corresponding to one of the properties appearing in the query. Then traversing the way of c until the root and adding to the query the tops met.

- Reformulation by specialization: consists also in locating the top c corresponding to one of the properties appearing in the query. But this time traversing under T which has like root c and then extracting all the tops, sheets from under tree and adding them to the query.

## IV.   FORMAL CONCEPT ANALYSIS

Among the mathematical theories recently found with important applications in computer science, lattice theory has a specific place for data organization, information engineering, and data mining. It may be considered as the mathematical tool that unifies data and knowledge or information retrieval [2, 3, 10, 11, 12].

### A.   Formal Context

A context is a triplet (G, M, I) which G and M are units and I ⊆ G×M is a relation ℜ. The elements of G are called the objects, M a finite set of elements called properties and R is a binary relation defined between G and M. The notation gIm means that "formal object g verifies property m in relation R".

Example: Let G = {s1, s2, s3, s4}, be a set of source and M = {p1, p2, p3, p4, p5} be a set of the properties (table 1). The mathematical structure which is used to describe formally this table is called a formal context (or briefly a context) [4, 9, 10, 13].

*B. Galois (Concept) lattice*

The family of the entire formal context ordinate by the relation $\geq$ is called Galois (or concepts) lattice. It consists in associating properties with sources, and organizing the sources according to these properties. Each such pair (G, M) is called a formal concept (or briefly a concept) of the given context. The set G is called the extent, the set M the intent of the concept (G, M). Between the concepts of a given context there is a natural hierarchical order, the "subconcept-superconcept" relation. In general, a concept c is a subconcept of a concept d (and d is called a superconcept of c) if the extent of c is a subset of the extent of d (or equivalently: if the intent of c is a superset of the intent of d). An efficient algorithm for extracting the set of all concepts of a given context is Ganter's `Next Closure' algorithm [11]. The algorithm can efficiently compute all concepts C (G, M, I) from a context (G,M,I). The concept lattice corresponds to the formal context of table 1 is presented in figure 2 (Hasse diagram). A line diagram consists of circles, lines names of all objects and all attributes of the given context. The circles represent the concepts.

TABLE I.    AN EXAMPLE OF FORMAL CONTEXT

| G×M | p1 | P2 | p3 | p4 | p5 |
|-----|----|----|----|----|----|
| s1  | 1  | 1  | 0  | 1  | 0  |
| s2  | 0  | 0  | 1  | 0  | 1  |
| s3  | 1  | 1  | 1  | 0  | 1  |
| s4  | 1  | 1  | 1  | 1  | 0  |

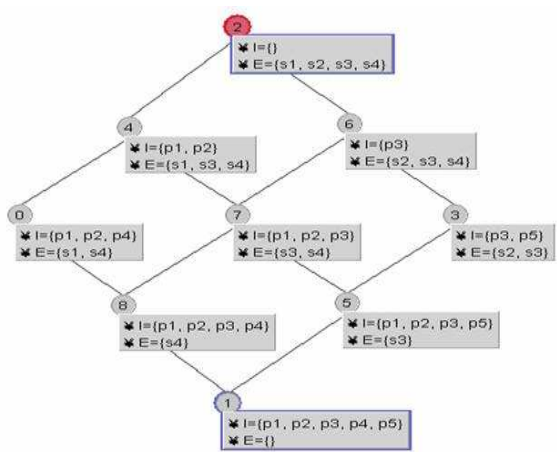

Figure 2.   Galois[2] lattice corresponding to formal context (table 1)

The lattice provides a visualization of the concept relationships that are implicit in the data. The attributes p1, p2, and p3 describe a subconcept of the concept of the propriety p3. The extent of this subconcept consists of the properties p1, and p2.

---

V.    IMPLEMENTATION AND RESULTS

*A. Building concepts lattice*

(i) Data sources

The data used for test were extracted from titles of a subset of documents from the LRIT (Research laboratory in computer science and Telecommunications). The laboratory LRIT consists of several research groups, their activities are based on the interaction between the contents (audio, image, text, video. Consequently there are several publications in several fields (image processing, signal processing, data mining, data engineering, information retrieval …), it there has also other publications heterogeneous, which requires a system that makes it possible to determine the interconnections between work of the various members to make emerge and understand the orientations of principal research in the team and also in laboratory LRIT, and thus to provide explanations on the research task.

For efficient purposes, the data that was extracted from the documents were stored in XML database file; which is used for the extraction of the properties or to post the results with the users. Each publication (or source) is represented by two tags <document...> and </document>. It has an attribute number with value 1 and two child elements author and title, there is also the possibility to add of extra information concerning the publications in these XML file. Figure 3 shows the listing of document 1, 2 and 3 from data set. Each document in the collection has a corresponding title, author's, and but not necessarily an abstract.

```
<?xml version="1.0" encoding="UTF-8"?>
<documents>
    <document nom="dcument_1">
        <auteur>Amine A</auteur>
        <auteur>Elakadi A></auteur>
        <auteur>Rziza M</auteur>
        <auteur>Aboutajdine D</auteur>
        <title>ga-svm and mutual information based frequency feature selection for
face recognition</titre>
    </document>
    <document nom="dcument_2">
        <auteur>El Fkihi S</auteur>
        <auteur>Daoudi M></auteur>
        <auteur>Aboutajdine D</auteur>
        <title>the mixture of k-optimal-spanning-trees based probability
approximation: application to skin detection image and vision computing</titre>
    </document>
    <document nom="dcument_3">
        <auteur>El Hassouni M</auteur>
        <auteur>Cherifi H></auteur>
        <auteur>Aboutajdine D</auteur>
        <title>hos-based image sequence noise renoval</titre>
    </document>
```

Figure 3.   XML file

Document term frequency was computed for each term extracted after applying the following techniques from the "classic blueprint for automatic indexing" [14]:

−    Segmentation: this is the process of selecting distinct terms from the individual documents in the collection. For our implementations, we broke hyphenated terms into their constituents, as well as ignoring punctuation and case.

−    Stop wording: this is the process of removing frequently occurring terms such as 'is', and 'of' which make little discrimination between documents.

(ii) Lattice construction

The problem of calculation of the concepts lattice from a formal context made object of several research tasks. Many algorithms have been proposed for generating the Galois lattice from a binary relation [2, 4, 9, 13, 15, 16]. A comparison of the performances of the algorithms proposed for the generation of the lattices and their corresponding Hasse diagrams are presented in algorithm Add Intent [10]. Among the algorithms proposed, some have specificity to perform an incremental building of concepts lattices starting from formal contexts [2, 4, 10]. This aspect is particularly interesting for the application of concepts lattices to our problem of research in the publications of the LRIT. Indeed, the queries users can be inserted in the lattice representing the documents (or publications). Following this insertion, it is possible to determine the most relevant documents guarantors with the criteria expressed by user in his query.

Our procedure for implementation FCA's, concept lattice involves three stages : constructing a matrix of document-term relations using data stored from XML file ; concept extraction using Add Intent algorithm; and partial ordering of formal concepts. The resulting internal data structure is ten written out to a file where it may be later used for querying (figure 4).

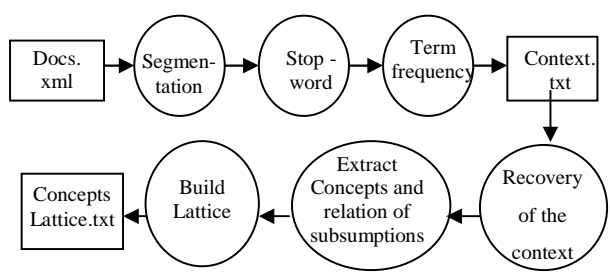

Figure 4.    Process of building of concepts lattice

(iii) Query Insertion

Our idea is to consider the user query as a new source whose properties are the terms of the query. This source will be added to the lattice Li produced by I first objects of the context in ways incremental using the algorithm Add Intent [10]. This addition will transform the lattice Li; new nodes will be added and others will be modified. It is necessary all the same to notice the appearance of a new concept which has like intention exactly the terms of the query.

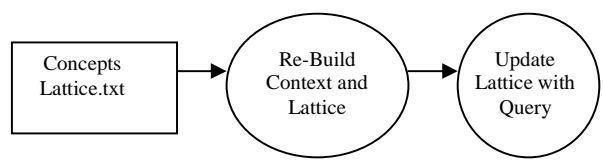

Figure 5.    Query Insertion

(iv) Document Ranking

Documents were ranked based on the number of edges away from the concept in which the query had mapped in the augmented lattice. Formal concepts in the lattice were considered for ranking only when its attribute set intersects with those of the query and that it is neither the supremum nor the infimum. Documents that were equally distant from the query would achieve the same rank. The lattice traversal implementation was simply done using a breadth-first search.

### B.    Discussion of Results

−    In first let us assume that we have an example of context for 5 documents and 6 properties (table 2). The lattice corresponding is presented in the figure 6.

TABLE II.    An example of formal context from database source

| M\G | d1 | d2 | d3 | d4 | d5 |
|---|---|---|---|---|---|
| image | 1 | 1 | 1 | 0 | 0 |
| detection | 0 | 0 | 0 | 1 | 1 |
| Segmentation | 1 | 1 | 0 | 1 | 0 |
| Classification | 0 | 0 | 1 | 0 | 0 |
| vision | 0 | 0 | 0 | 0 | 1 |
| probability | 1 | 0 | 0 | 1 | 0 |

The Galois lattice establishes a clustering of the data sources. Each formal concept of the lattice represents in fact a class. For example, the concept ({d1, d4}, {probability, segmentation}) puts in the same class the data sources d1 and d4. These two sources are in this class because they are the only ones to share the properties probability, segmentation. The lattice establishes also a hierarchy between the classes. One can read that there formal concept is ({d1, d4}, {probability, segmentation}) more particular than ({d1, d2, d4}, {segmentation}) in the direction where it has more properties it is what results in the fact that {d1, d2} included {d1, d2, d4}. We note that this hierarchy is with double directions, i.e., the lattice is a kind of "tree structure" with two "roots": ({d1, d2, d3, d4, d5}, {}) and ({}, {classification, detection, image, probability, segmentation, vision}) that we will respectively call top and bottom. Displacement towards top corresponds to generalization and that towards bottom to specialization.

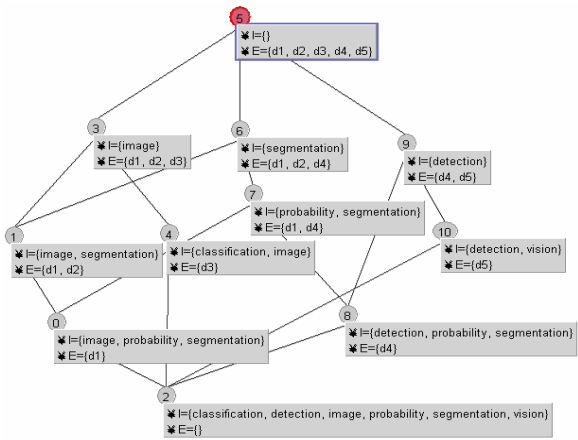

Figure 6. Trellis of concepts associated to table 2

Let us suppose that the user formulates the following query: detection, segmentation. This query will be represented as follows: ({Query}, {detection, segmentation}). It can be connected with a new query source which has the properties: detection, segmentation. And the lattice, after the addition of the query, will change as it is illustrated in the figure 7.

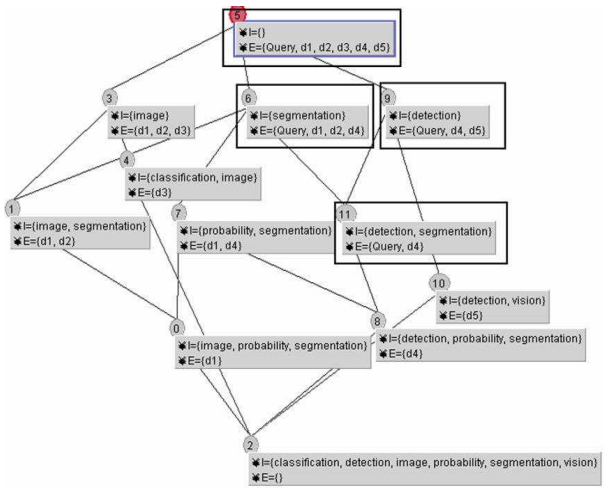

Figure 7. Trellis of concepts after query insertion

In our example (figure 7), the user query will generate the following answers: On the first level of the node ({Query}, {detection, segmentation}), the extension part comprises the d4 source. What means that d4 is the most relevant source for this query in the lattice and thus one will attribute him the rank=0. On the second level, the provided answers will be the d1 sources, d2, and d5 and attributes their consequently the rank=1. The d1 sources and d2 in common have with the query the property `segmentation', whereas the source d5 division the property `detection' with the query. And thus the result will be presented like continuation:

0 - d4

1 - d1

1 - d2

1 - d5

It is noticed that, indeed, the turned over sources are all relevant in the direction where they have at least a property desired by the user and which they were provided in decreasing order relevance.

-   In second step, we built the concept lattice (figure 8) based on a formal context of 7 documents containing the properties used in ontology presented in figure 1.

In query insertion, let us suppose that we formulate the following query: `detection of contour' (shortened by dc.). This query will be represented as follows:

P= ({Query}, {dc.}). It can be connected with a new Query source which has the properties: `detection of contour'. The lattice, after addition of the query, will change like below figure 9.

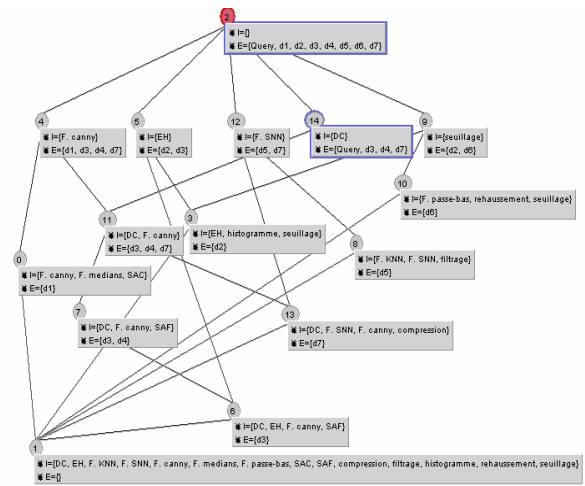

Figure 8. Concept lattice associated to 7 documents containing the concepts used in figure 1

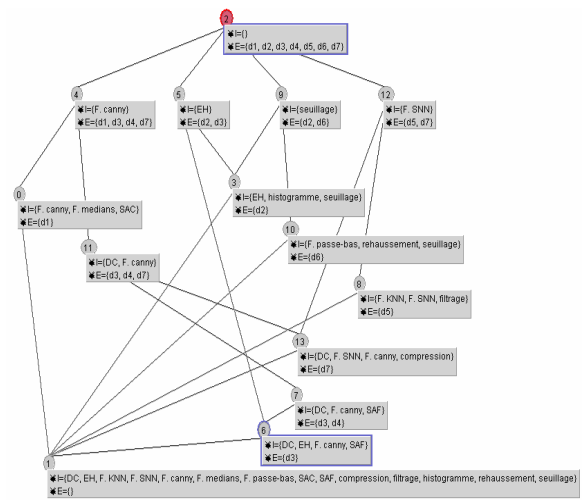

Figure 9. Concept lattice after query insertion

In our example (figure 9), the user query will generate the following answers:

On the first level of the node ({Query, d3, d4, d7}, {dc.}), the extension part comprises the d3 sources, d4, and d7. What means that the latter are the most relevant sources for this query in the lattice and thus one their will attribute their rank=0. On the second level there is the top concept consequently it is necessary to stop. The result will be presented like continuation:

0-d3

0-d4

0-d7

On the other hand, using the semantic relations between the properties in ontology in field of image processing (figure 2). The query reformulation by specialization gives: {dc. Canny filter}. After the insertion of this new query in the lattice (figure 10) the result turning over is as follows:

0-d3

0-d4

0-d7

1-d1

Query reformulation by generalization gives: {dc. segmentation by approach (border) (shortened by SAF), segmentation}. After the insertion of this one in the lattice one has like result:

0-d3

0-d4

1-d7

The result did not change because the properties of the new query (after reformulation by generalization) division the same sources (figure 11).

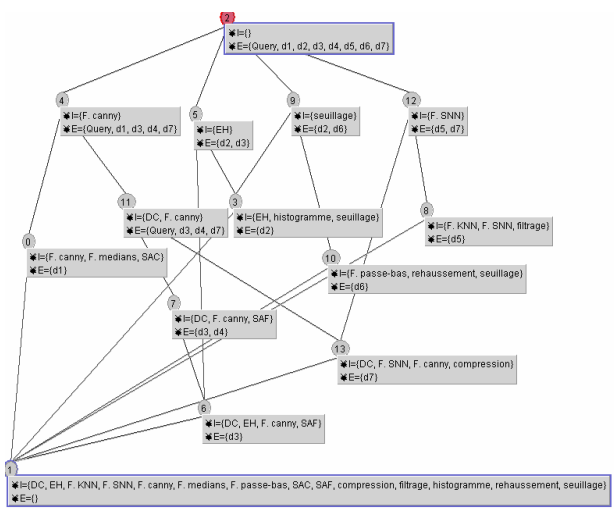

Figure 10. Concept lattice after query reformulation by specialization

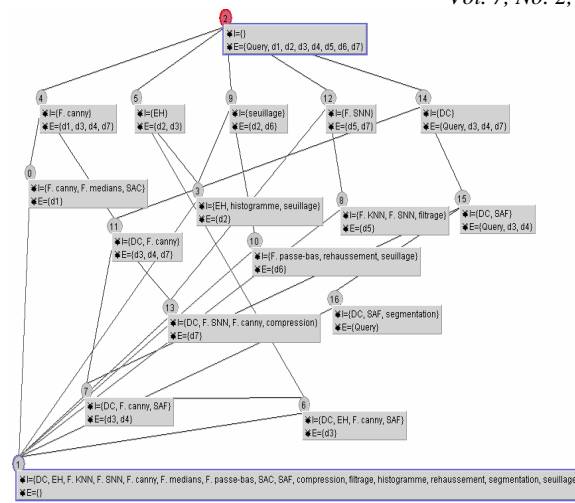

Figure 11. Concept lattice after query reformulation by generalization

We saw that this ontology enables us to take into account the semantic relations between the properties. Moreover, the possibility of making a research by specialization or generalization has an advantage of having more relevant sources to add to the initial result. The choice of reformulation depends on the user. It is a reformulation by generalization, the added source can be very general and consequently not very precise compared to what is wished by user. And it is a reformulation by specialization; the added source can cover with many details only one small portion of what user asks. But in no case an added source cannot be completely isolated compared to what is wished by user.

## VI. CONCLUSION

We presented in this paper an proposal in the Information Retrieval (IR), using Formal Concepts Analysis (FCA). The concept lattice evolves during the process of IR; the user is not more restricted with a static structure calculated once for all, and the system is domain independent and operates without resorting to thesauruses or other predefined sets of indexing terms. The system implemented allows the user to navigate in hierarchy of concepts to research the relevance documents to his query. To perform the IR we established ontology in field of image processing that enables us to take into account the semantic relations between the properties. Moreover, we also improved the results by using two steps for query reformulation, reformulation by generalization, and by specialization, which show the more relevant documents returned by system to the user query.